\documentclass[12pt]{article}

\usepackage{array} 
\usepackage{epsfig}
\usepackage{amssymb}
\usepackage{graphics,graphpap}

\renewcommand{\thefootnote}{\fnsymbol{footnote}}

\begin{document}

\begin{titlepage}
\begin{flushright}
\begin{tabular}{l}
IPPP/06/61\\
DCPT/06/122
\end{tabular}
\end{flushright}
\vskip1.5cm
\begin{center}
   {\Large \bf\boldmath
 Time-dependent CP Asymmetry in $B\to K^*\gamma$\\[7pt]
    as a (Quasi) Null Test of the Standard Model}
    \vskip1.3cm {\sc
Patricia Ball\footnote{Patricia.Ball@durham.ac.uk} and 
    Roman Zwicky\footnote{Roman.Zwicky@durham.ac.uk}}
  \vskip0.5cm
        {\em IPPP, Department of Physics, 
University of Durham, Durham DH1 3LE, UK} 

\vskip2cm


\vskip3cm

{\large\bf Abstract:\\[10pt]} \parbox[t]{\textwidth}{
We calculate the dominant Standard Model contributions to the time-dependent CP
asymmetry in $B^0\to K^{*0}\gamma$, which is $O(1/m_b)$ in QCD
factorisation. We find that, including all relevant hadronic effects,
in particular from soft gluons, 
the asymmetry $S$ is very small, $S=-0.022\pm 0.015^{+0}_{-0.01}$,
and smaller than suggested recently from 
dimensional arguments in a $1/m_b$ expansion. Our result
implies that any significant deviation of the asymmetry from zero, and
in particular a confirmation of the current experimental central
value, $S_{\rm HFAG}=-0.28\pm 0.26$, 
would constitute a clean signal for new
physics.
}

\vfill

\end{center}
\end{titlepage}

\setcounter{footnote}{0}
\renewcommand{\thefootnote}{\arabic{footnote}}

\section{Introduction}

The radiative decay $b\to s\gamma$ has been extensively studied as a
probe of both the flavour structure of the Standard Model (SM) and
new physics beyond the SM (see Ref.~\cite{hurth} for a review). While
the vast majority of studies has focused on the prediction of the decay
rate for exclusive and both spectra and decay rate for inclusive $b\to
s\gamma$ decays, there is one rather peculiar feature of this
process which has attracted far less attention, namely that, in the
SM, the emitted photon is predominantly left-handed in $b$, and
right-handed in $\bar b$ decays. This  is due to the fact
that, in the language of effective field theories, the dominant 
contribution is from the chiral-odd
dipole operator $\bar s_{L(R)} \sigma_{\mu\nu} b_{R(L)}$. 
As only left-handed quarks
participate in the weak interaction, an effective operator of this
type necessitates, in the SM, a
helicity flip on one of the external quark lines, which results in a factor
$m_b$ (and a left-handed photon) for $b_R\to s_L\gamma_L$ 
and a factor $m_s$ (and a right-handed photon) 
 for $b_L\to s_R\gamma_R$. Hence, the emission of
right-handed photons is suppressed by
roughly a factor $m_s/m_b$. This suppression
 can easily be alleviated in a large number
of new physics scenarios where the helicity flip occurs on an
internal line, resulting in a factor $m_i/m_b$ instead of $m_s/m_b$. 
A prime example are left-right symmetric models
\cite{LRS}, whose impact on the photon polarisation was discussed in
Ref.~\cite{gronau}. These models also come in a supersymmetric version 
whose effect on $b\to s\gamma$ was investigated in Ref.~\cite{frank}. 
Supersymmetry with no left-right symmetry can also provide large contributions
to $b\to s\gamma_R$, see Ref.~\cite{susy} for recent studies. Other
potential sources of large effects which have been studied 
are warped extra dimensions
\cite{warped} or anomalous right-handed top
couplings \cite{anomalous}. Unless the amplitude for $b\to s \gamma_R$
is of the same order as the SM prediction for $b\to s \gamma_L$, or the
enhancement of $b\to s \gamma_R$ goes along with a suppression of 
$b\to s \gamma_L$, the impact
on the branching ratio is small, as the two helicity
amplitudes add incoherently. This implies there can be a
substantial contribution of new physics to $b\to s\gamma$ 
escaping detection when only branching ratios are measured.

Although the photon helicity is, in principle, an observable, it is
very difficult to measure directly. It can, however, be accessed
indirectly, for instance in the time-dependent CP asymmetry in $B^0\to
K^{*0}\gamma$, which relies on the interference of both left and right
helicity amplitudes and vanishes if one of them is absent. This method
was first suggested in Ref.~\cite{gronau} and later discussed in more
detail in Refs.~\cite{grin04,grin05}. It is rather special in the
sense that usually new physics modifies the SM predictions for
time-dependent CP asymmetries by affecting the mixing phase (as in
$B_s\to J/\psi \phi$, see for instance Ref.~\cite{BF06}), introducing
new weak phases or moderately changing the size of the decay amplitudes which,
in the absence of precise calculational tools, makes it very hard to
trace its impact. In contrast, the time-dependent CP asymmetry in
$B^0\to K^{*0}\gamma$ is very small in the SM, irrespectively of
hadronic uncertainties, by virtue of the helicity suppression of one decay
amplitude, and new physics enters by relieving that
suppression. The smallness of the asymmetry in the SM, and the possibility of
large effects from new physics, makes it one of the prime candidates
for a so-called ``null test'' of the SM, as recently advertised in 
Ref.~\cite{null}.

Other channels and methods to
probe the photon helicity have been discussed in Refs.~\cite{other,soni}. In
this letter, however,  we focus on the time-dependent CP asymmetry in $B^0\to
K^{*0}\gamma$. It is given by 
\begin{equation}\label{-1}
A_{CP} = \frac{\Gamma(\bar B^0(t)\to \bar K^{*0}\gamma) -
               \Gamma(     B^0(t)\to      K^{*0}\gamma)}{
               \Gamma(\bar B^0(t)\to \bar K^{*0}\gamma) +
               \Gamma(     B^0(t)\to      K^{*0}\gamma)}
= S \sin(\Delta m_B t ) - C \cos(\Delta m_B t)\,,
\end{equation}
where $K^{*0}$ and $\bar K^{*0}$ are observed via their decay into the CP
eigenstate $K_S\pi^0$.
The term involving an interference of photons with different
polarisation is $S$, for which the following experimental results are
available from the $B$ factories:
\begin{equation}
\begin{array}[b]{lll@{\hspace*{15pt}}l}
S_{\rm BaBar} & = -0.21 \pm 0.40\,({\rm stat}) \pm 0.05\, ({\rm syst})
& \quad \mbox{BaBar \cite{BaBar}} 
& \mbox{($232\cdot 10^6$ $B\bar B$ pairs),} 
\\[5pt]
S_{\rm Belle} & =  -0.32^{+ 0.36}_{-0.33}\,({\rm stat}) \pm 0.05\, ({\rm syst})
& \quad \mbox{Belle \cite{Belle}} 
&  \mbox{($535\cdot 10^6$ $B\bar B$ pairs),} 
\end{array}
\end{equation}
with the HFAG average $S_{\rm HFAG} = -0.28\pm 0.26$ \cite{HFAG}. While these
results are  compatible with zero at the 1$\sigma$ level, the
central values of both BaBar and Belle are in agreement and
interestingly large.
A drastic reduction of the
experimental uncertainty will probably be difficult at the LHC, but 
can be achieved at a Super $B$ factory, with
an anticipated statistical uncertainty of $S$ of 0.07 with 10~ab$^{-1}$
of data \cite{superB} and 0.04 with 50~ab$^{-1}$ \cite{superKEK}.

\begin{figure}
$$\epsfxsize=0.2\textwidth\epsffile{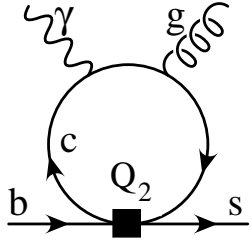}$$
\caption[]{\small Dominant contribution to $b\to s\gamma g$. A second diagram
with photon and gluon vertices exchanged is implied.}\label{fig1}
\end{figure}

In order to clearly distinguish any new physics signal from the SM
background, one needs to know the latter as precisely as possible. As
discussed above, one
contribution comes from $b_L\to s_R\gamma_R$, with 
a helicity flip on the $s$ quark line; it generates the
contribution
\begin{equation}\label{0}
S^{{\rm SM},s_R} = - \sin(2\beta)\,\frac{m_s}{m_b}\left(2 + O(\alpha_s)\right)
\end{equation}
to the CP asymmetry, with $\beta$ being one of the angles of the CKM unitarity
triangle. At leading order in $\alpha_s$, $S^{{\rm SM},s_R}$ is free from
hadronic uncertainties. As  pointed out in Ref.~\cite{grin04},
another mechanism to remove the helicity suppression of $b\to s \gamma_R$ is
to emit an additional gluon. The dominant
contribution to this mechanism
is via a $c$-quark loop and is shown in Fig.~\ref{fig1}.
In inclusive decays this is a bremsstrahlung correction and
can be calculated in perturbation theory \cite{grin04}. 
In exclusive decays, on the
other hand, the gluon can be either hard or soft. If it is hard,
it attaches to the spectator quark, which
induces $O(\alpha_s)$ corrections to (\ref{0}). If it is soft, it 
has to be interpreted as a parton in one of the external 
hadrons. Stated differently, if the gluon is soft, the amplitude  
involves higher Fock states of the
$B$ and $K^*$. 
A data-driven method to distinguish this contribution from that of the
dipole operator $\bar s_{L(R)} \sigma_{\mu\nu} b_{R(L)}$ was discussed
in Ref.~\cite{soni} and relies on the Dalitz-plot analysis of decays
of type $B^0\to \gamma K_S+\,$neutrals, where neutrals stands for
$\pi^0$, $\eta^{(')}$, $K_S$, light vector mesons or any combination
of these particles.
In Ref.~\cite{grin04} it was shown, in the framework of
soft-collinear effective theory (SCET), that contributions from $b\to
s\gamma g$ are suppressed by
one power of $m_b$ with respect to the left-handed photon emission,
which confirms the results obtained in QCD factorisation
\cite{QCDfac,BoBu}, where an explicit $O(\alpha_s)$ calculation
demonstrated, to leading order in $1/m_b$, the absence of 
right-handed photons in $\bar B^0\to \bar K^{*0}\gamma$. A SCET
analysis of the CP asymmetry in Ref.~\cite{grin05} estimated the
size of the $1/m_b$ corrections to $S^{\rm SM}$ induced
by $b\to s\gamma g$ as $\sim \pm 0.1$, but is based on
dimensional counting of the operators involved rather than a
calculation of the relevant matrix elements. Another
calculation, in perturbative QCD, gives $S^{\rm SM}_{\rm pQCD} 
= -(3.5\pm 1.7)\%$
\cite{sanda}, including effects mainly from hard gluons; 
the contribution of soft gluons is treated in a model-dependent way. 

The purpose of this letter is to provide a calculation of the 
soft-gluon contribution to the time-dependent CP asymmetry in $B\to
K^*\gamma$ induced by the $c$-quark loop shown in
Fig.~\ref{fig1}. The method we use are QCD sum rules on the
light-cone. It turns out that the relevant hadronic parameters were
 calculated already previously, in 1997, in Ref.~\cite{alex},
using the method of local QCD sum rules. The motivation at the time
was to estimate long-distance corrections to the branching ratio of
$B\to K^*\gamma$. In fact those corrections were first discovered 
for the inclusive process \cite{misha}.
In this letter, we show that the same parameters
also enter the time-dependent  CP asymmetry in $B\to
K^*\gamma$ and present a new calculation of their values.

\section{The CP Asymmetry}

Let us define the amplitudes of the decay of $B$ mesons into $K^*$ and
left- or right-handed photons in the following way:
\begin{equation}
\bar{\cal A}_{L(R)} = {\cal A}(\bar B^0\to \bar K^{*0}
\gamma_{L(R)})\,,  \qquad
{\cal A}_{L(R)} = {\cal A}(B^0\to K^{*0}
\gamma_{L(R)})\,.
\end{equation}
Neglecting, as usual, the small width difference between $B^0$ and
$\bar B^0$, the time-dependent CP asymmetry is then given by (\ref{-1})
with
\begin{eqnarray}
S & = & \frac{2 \,{\rm Im}\,\left(\frac{q}{p}({\cal A}_L^* \bar{\cal A}_L + 
                                       {\cal A}_R^* \bar{\cal A}_R)\right)}{
        |{\cal A}_L|^2 + |{\cal A}_R|^2 + |\bar{\cal A}_L|^2 + |\bar{\cal
                                       A}_R|^2}\,,
\qquad
C  =   \frac{|{\cal A}_L|^2 + |{\cal A}_R|^2 - |\bar{\cal A}_L|^2 - 
               |\bar{\cal A}_R|^2}{
        |{\cal A}_L|^2 + |{\cal A}_R|^2 + |\bar{\cal A}_L|^2 + |\bar{\cal
                                       A}_R|^2}\,.\label{5}
\end{eqnarray}
Here $q/p$ is given in terms of the $B^0$-$\bar B^0$ 
mixing matrix $M_{12}$, in 
the standard convention for the parametrisation of the CKM matrix, by
$$
\frac{q}{p} = \sqrt{\frac{M_{12}^{*}}{M_{12}}} = e^{-2 i \beta}\,.
$$

Extending in an obvious way
the notations introduced in Ref.~\cite{BoBu} in the context
of QCD factorisation, the
decay amplitudes can be written as
\begin{eqnarray}
\bar{\cal A}_{L(R)} &=& \frac{G_F}{\sqrt{2}}\,\left( \lambda_u
a_{7}^u(\bar K^*\gamma_{L(R)}) +
\lambda_c a_{7}^c(\bar K^*\gamma_{L(R)})\right) \langle \bar K^*
\gamma_{L(R)} | Q_7^{L(R)} | \bar
B\rangle
\nonumber\\
&\equiv& \frac{G_F}{\sqrt{2}}\,\left( \lambda_u
a_{7L(R)}^u + \lambda_c a_{7L(R)}^c\right) \langle \bar K^*
\gamma_{L(R)} | Q_7^{L(R)} | \bar B\rangle\,.\label{ME}
\end{eqnarray}
In QCD factorisation, $a_{7L}^{c,u}$ are of order 1 in a $1/m_b$
expansion \cite{BoBu},\footnote{The
  $a_7^{c,u}$ calculated in Ref.~\cite{BoBu}, to leading order in
  $1/m_b$, coincide with our
  $a_{7L}^{c,u}$, whereas $a_{7R}^{c,u}$ are set zero in \cite{BoBu}. Our
expression (\ref{ME}) is purely formal and does not imply that
$a_{7R(L)}^{c,u}$ factorise at order $1/m_b$. As a matter of fact,
they don't.}
\begin{equation}
a_{7L}^{c,u} = C_7 + O(\alpha_s,1/m_b)\,,
\end{equation}
with $C_7$ being the Wilson coefficient of the operator $Q_7$. The
complete set of operators and formulas for the Wilson coefficients can
be found in Ref.~\cite{misiak}.
$a_{7R}^{c,u}$, on the other hand, are of order $1/m_b$ \cite{grin04,grin05}. 
$\lambda_p = V_{ps}^* V_{pb}$ and the operators $Q_7^{L(R)}$ are given by
$$
Q_7^{L(R)} = \frac{e}{8\pi^2}\, m_b \bar s \sigma_{\mu\nu} 
             \left(1 \pm \gamma_5\right)b F^{\mu\nu}\,;
$$
$Q_7^{L(R)}$ generates left- (right-) handed photons in the decay
$b\to s\gamma$. 
The matrix element in (\ref{ME}) can be expressed in terms of the
	     form factor $T_1^{B\to K^*}$  as
\begin{eqnarray}
\lefteqn{\langle \bar K^*(p,\eta) \gamma_{L(R)}(q,e) | Q_7^{L(R)} | \bar
B \rangle =}\hspace*{1cm}\nonumber\\
&=& -\frac{e}{2\pi^2}\, m_b T_1^{B\to K^*}(0) \left[
\epsilon^{\mu\nu\rho\sigma} e_\mu^* \eta_\nu^* p_\rho q_\sigma \pm i
\{ (e^* \eta^*) (pq) - (e^*p)(\eta^* q)\}\right]
\nonumber\\
&\equiv& -\frac{e}{2\pi^2}\, m_b T_1^{B\to K^*}(0) S_{L(R)}\,,\label{6}
\end{eqnarray}
where $S_{L,R}$ are the helicity amplitudes corresponding to left- and
right-handed photons, respectively, and $e_\mu(\eta_\mu)$ is the polarisation
four-vector of the photon $(K^*)$. The definition of $T_1^{B\to K^*}$
can be found in Ref.~\cite{BZ04}, and an updated value in Ref.~\cite{proc}.
In the Wolfenstein parametrisation of the CKM matrix, $\lambda_u\sim
\lambda^4$ and is doubly Cabibbo suppressed with respect to $\lambda_c\sim
\lambda^2$, so we drop this contribution from now on. With 
$\lambda_u$ set to zero, the direct CP asymmetry $C$ 
in (\ref{5}) vanishes. 

\section{\boldmath Calculation of $a_{7R}^c$ in the SM}

One contribution to $\bar{\cal A}_R$ is very well known and comes from
the $m_s$ dependent part of the full electromagnetic dipole operator
$Q_7$,
\begin{equation}\label{Q7}
Q_7 = \frac{e}{8\pi^2} \left[ m_b \bar s \sigma_{\mu\nu} (1+\gamma_5)
b + m_s \bar s \sigma_{\mu\nu} (1-\gamma_5) b\right] F^{\mu\nu} \equiv Q_7^L
+ \frac{m_s}{m_b}\, Q_7^R\,.
\end{equation}
Hence, $a_{7R}^c$ is given by 
\begin{equation}
a_{7R}^c = \frac{m_s}{m_b}\, C_7 + 
O\left(\frac{1}{m_b}\,,\frac{\alpha_s}{m_b}\right).
\end{equation}
As discussed above, all contributions to $\bar{\cal A}_R$ must include 
a helicity flip of the
$s$ quark, which in the above is done by including the effects from a
non-vanishing $s$ quark mass. Another possibility to relieve the
helicity suppression of right-handed photons is by considering, at
parton level, a three-particle final state with an additional
gluon. The dominant contribution (with the largest Wilson coefficient)
to this process comes from the
operator 
$$Q_2^c = [\bar s\gamma_\mu (1-\gamma_5) c][\bar
c\gamma^\mu(1-\gamma_5)b]
$$
and is shown in Fig.~\ref{fig1}. As  the $c$ quark has sufficiently
large virtuality in the loop (the photon is
on-shell and the gluon nearly so), 
the diagram is dominated by short distances and
can be expanded in inverse powers of $m_c$. To do so, we follow 
Ref.~\cite{alex} and rewrite
$Q_2^c$ as
\begin{equation}\label{10}
Q_2^c = \frac{1}{3}\,[\bar c \gamma_\mu (1-\gamma_5) c][\bar
s\gamma^\mu(1-\gamma_5)b] + 2 [\bar s\gamma_\mu (1-\gamma_5)\, 
\frac{\lambda^a}{2}\,c][\bar
c\gamma^\mu(1-\gamma_5)\,\frac{\lambda^a}{2}\,b]\,.
\end{equation}
Confirming the result of Ref.~\cite{alex}, we find that 
the short-distance expansion of the diagram in Fig.~\ref{fig1}
yields
\begin{eqnarray}
Q_F &=& i e^{*\mu} \int\, d^4 x e^{iqx}\, {\rm T}
\left\{[\bar c(x) \gamma_\mu c(x)]\, Q_2^c(0)\right\}\nonumber\\
&=&  -\frac{1}{48\pi^2 m_c^2} (D^\rho F^{\alpha\beta}) [\bar s
  \gamma_\rho (1-\gamma_5) g\widetilde G^a_{\alpha\beta}
  \frac{\lambda^a}{2}\, b] + \dots\label{sd}
\end{eqnarray} 
where $F^{\alpha\beta} = i (q^\alpha e^{*\beta} - q^\beta
e^{*\alpha})$ corresponds to an outgoing photon and the dots denote
terms of higher order in $1/m_c$. Note that the contribution of the
first term in (\ref{10}) vanishes for an on-shell photon. The
contribution of $Q_F$ to the decay amplitude is
$$
{\cal A}_{Q_F}(\bar B\to \bar K^{*} \gamma) =
-\frac{2e}{3}\,\langle \bar K^{*}\gamma | Q_F | \bar B\rangle\,,
$$
where $2/3$ is the electric charge of the $c$ quark and the minus sign
comes from the EM interaction operator.
At this point we
would also like to make explicit our conventions for the strong and
electromagnetic couplings. We use the covariant derivative
$$D_\mu = \partial_\mu + i e Q_f B_\mu - i g A_\mu^a
\,\frac{\lambda^a}{2}$$
for a fermion with electric charge $Q_f$. Here
$e=+\sqrt{4\pi\alpha}$ which is consistent with the sign-convention for $Q_7$,
Eq.~(\ref{Q7}).\footnote{The sign of the strong coupling $g$ differs
  with respect to Ref.~\cite{BoBu}, which however does not matter
  as all final expressions contain only factors $g^2$.}
The contribution of $Q_2^c$ to $a_{7R}^c$ is hence governed by the
matrix element $\langle K^*\gamma | (D^\rho F^{\alpha\beta}) [\bar s
  \gamma_\rho (1-\gamma_5) g\widetilde G^a_{\alpha\beta}
  \frac{\lambda^a}{2}\, b] | B\rangle$, which, again following
Ref.~\cite{alex}, can be parametrised as
\begin{eqnarray}
\lefteqn{\langle \bar K^*(p,\eta)\gamma(q,e) | (D^\rho F^{\alpha\beta}) [\bar s
  \gamma_\rho (1-\gamma_5) g\widetilde G^a_{\alpha\beta}
  \frac{\lambda^a}{2}\, b] | \bar B(p+q)\rangle=}\hspace*{1.5cm}
\nonumber\\
&=& 2 \langle \bar K^*(p,\eta) | \bar s \gamma_\mu q^\mu (1-\gamma_5) g
  \widetilde G_{\alpha\beta} b | \bar B(p+q)\rangle  e^{*\alpha} q^\beta
\nonumber\\
&=&2 \left\{L \epsilon_{\mu\nu\rho\sigma} e^{*\mu} \eta^{*\nu} p^\rho q^\sigma
  + i \widetilde L [ (e^{*} \eta^*)(pq) - (e^* p)(\eta^*
  q)]\right\}\nonumber\\
& = & (L+\widetilde L) S_L + (L-\widetilde L) S_R\,,
\end{eqnarray}
where $S_{L,R}$ are the photon helicity structures defined in
(\ref{6}). The operator $Q_2^c$ thus induces power corrections of type
$(L\pm \widetilde L)/(m_c^2 m_b)$ to $a_{7L}^c$ and $a_{7R}^c$, respectively.
As already mentioned before, these power corrections have previously
been considered in Ref.~\cite{alex}. Before we
present a new calculation of $L$ and $\widetilde L$, let us  
finally give their contribution to
$a_{7R}^c$:
\begin{equation}\label{comp}
a_{7R}^c = C_7 \,\frac{m_s}{m_b} - C_2\, \frac{L-\widetilde L}{36
  m_c^2 m_b T_1^{B\to K^*}(0)}\,.
\end{equation}
Corrections to this expression are of order $\alpha_s/m_b$, come with
smaller (penguin) Wilson coefficients or are of higher order in $1/m_{b,c}$.
What about the convergence of the $1/m_c$ expansion? For the
inclusive decay $b\to s\gamma$ this question was studied in Ref.~\cite{add}.
Higher terms in the short-distance expansion of (\ref{sd}) generate
operators with higher order derivatives acting on $F^{\alpha\beta}$,
generating powers of the photon momentum $q$,
and on $\widetilde{G}_{\alpha\beta}$, generating new hadronic matrix 
elements. A complete calculation of these additional
contributions to (\ref{comp}) is not possible with the presently
available methods, but we can try to give an estimate. 
As found in Ref.~\cite{add}, 
the expansion parameter of the $1/m_c$ expansion is  
$t=(q\cdot D)/(2m_c^2)$ with $D$ acting on the gluon field strength
tensor. The hadronic matrix elements with  additional powers of $D$
can be estimated as
\begin{equation}\label{est}
\langle K^* | \bar s\, D^n \widetilde G\, b| B\rangle \sim (\Lambda_{\rm
  QCD})^n \langle K^* | \bar s\, \widetilde G\, b| B\rangle\,,
\end{equation} 
and hence $t\sim (m_B/2) \Lambda_{\rm
  QCD}/(2m_c^2) \approx 0.2$. Using (\ref{est}), the $1/m_c$ series
  can be resummed and enhances the term in (\ref{sd}) by a factor  1.1
  for $t=0.2$, and 1.3 for $t=0.4$.\footnote{The enhancement factor is
  given by
  the function $\overline F(t,t)$ defined in the last reference of
  \cite{add}.}
  Although (\ref{est})
  is only a crude estimate of the true value of these matrix elements, 
  this result suggests that the $1/m_c$ expansion converges well. 
  We also would like to mention, as noted in \cite{add}, that besides 
  the derivative expansion in the  gluon field there are further 
  higher-twist contributions from e.g.\  two gluon fields.
  These contributions, however, are suppressed by additional powers of 
  $\Lambda_{\rm QCD}^2/(m_c^2)$ \cite{add}.
  We shall
  include the effect of truncating the $1/m_c$ expansion by
  doubling the theoretical error of our final result for the CP asymmetry.

\section{\boldmath Non-factorisable Soft Gluon Effects: $L$ and $\widetilde L$}

The following results for $L$ and $\widetilde L$ 
were obtained in 1997, 
in Ref.~\cite{alex}, using the method of local QCD sum rules and
neglecting the effects of $SU(3)$ breaking:
\begin{equation}\label{alex}
L = (0.55\pm 0.1)\,{\rm GeV}^3,\qquad 
 \widetilde L = (0.70\pm 0.1)\,{\rm GeV}^3\,.
\end{equation}
Since then, a number of
studies \cite{studies} have demonstrated that the appropriate method to
calculate $B$ decay form factors from QCD sum rules is to use QCD
sum rules on the light-cone \cite{BBK,LCSR}. In this paper, we cannot
give any account of the method itself, but refer to the relevant
literature. Suffice it to say that one of the main ingredients in the
method are light-cone distribution amplitudes (DAs) 
of two- and three-particle Fock
states of the  final-state meson. These have
been known for some time for $\rho$ mesons \cite{BBKT,rho}, but
complete expressions for $K^*$ mesons will become available 
only later in 2006 \cite{prep} (see
also \cite{BZ05,BZ06_1}). In this paper, we include the first
(preliminary) results of this ongoing study in our calculation of
$L$ and $\widetilde L$. The light-cone sum rules read:
\begin{eqnarray}
\lefteqn{\frac{f_B m_B^2}{m_b}\, L e^{-m_B^2/M^2} =  m_b^4
\int_0^{1-m_b^2/s_0} d\alpha_2 e^{-m_b^2/(\bar \alpha_2
  M^2)}}
\nonumber\\
&&\times \left\{
\frac{1}{2\bar\alpha_2^2} \int_0^{1-\alpha_2}d\alpha_1 \left[
  \left(\frac{m_{K^*}}{m_b} \right) f_{K^*}^\parallel
  \Phi^\parallel_{3;K^*}(\underline{\alpha}) +
  \left(\frac{m_{K^*}}{m_b}\right)^2 f_{K^*}^\perp \bar\alpha_2 \left(
  \Psi^\perp_{4;K^*}(\underline{\alpha}) + 
  \Phi^{\perp(1)}_{4;K^*}(\underline{\alpha}) \right)\right]\right.
\nonumber\\
&&{}\hspace*{0pt}\left.- \left(\frac{m_{K^*}}{m_b}\right)^2\!
   f_{K^*}^\perp\! \left[
   \frac{1}{4\bar\alpha_2^2} \, {\rm I}\,[\Phi^\perp_{3;K^*}+ 
   2 (\Phi^{\perp(3)}_{4;K^*} + \Phi^{\perp(4)}_{4;K^*})] +
   \frac{d}{d\alpha_2}\! \left(\frac{1}{2\bar\alpha_2}\, 
   {\rm I}\,[\Phi^{\perp(1)}_{4;K^*} +
     \Phi^{\perp(4)}_{4;K^*}]\right)\! \right] \!\right\}\!,\label{1}
\\
\lefteqn{\frac{f_B m_B^2}{m_b}\, \widetilde{L} e^{-m_B^2/M^2} =  m_b^4
\int_0^{1-m_b^2/s_0} d\alpha_2 e^{-m_b^2/(\bar \alpha_2
  M^2)}}
\nonumber\\
&&\times \left\{
\frac{1}{2\bar\alpha_2^2} \int_0^{1-\alpha_2}d\alpha_1 \left[
  \left(\frac{m_{K^*}}{m_b} \right) f_{K^*}^\parallel
  \widetilde\Phi^\parallel_{3;K^*}(\underline{\alpha}) -
  \left(\frac{m_{K^*}}{m_b}\right)^2 f_{K^*}^\perp \bar\alpha_2 \left(
  \widetilde\Psi^\perp_{4;K^*}(\underline{\alpha}) + 
  \Phi^{\perp(2)}_{4;K^*}(\underline{\alpha}) \right)\right]\right.
\nonumber\\
&&{}\hspace*{15pt}\left.+ \left(\frac{m_{K^*}}{m_b}\right)^2
   f_{K^*}^\perp 
   \frac{1}{4\bar\alpha_2^2} \, {\rm I}\,[\Phi^\perp_{3;K^*}- 
   2 (\Phi^{\perp(1)}_{4;K^*} + 2\Phi^{\perp(2)}_{4;K^*} +
   \Phi^{\perp(3)}_{4;K^*})]\right\}.\label{2}
\end{eqnarray}
The above expressions are accurate up to terms of order
$(m_{K^*}/m_b)^3$, which are of higher twist, and $O(\alpha_s)$ corrections.
Here $\Phi(\underline{\alpha})=\Phi(\alpha_1,\alpha_2)$ are
three-particle DAs of the $K^*$ 
of twist 3 or 4 (as indicated by the index). The (rather lengthy)
definition of these DAs is given in Ref.~\cite{prep}. The variable
$\alpha_1$ can
be interpreted as the longitudinal momentum fraction carried by the
quark in the meson, whereas $\alpha_2$ is the momentum fraction
carried  by the antiquark. 
${\rm I}\,[\Phi]$ is a functional acting on the DA
$\Phi(\underline{\alpha})$, which is
defined as
$${\rm I}\,[\Phi] = \int_0^{\alpha_2} dx \int_0^{1-x} d\alpha_1
\Phi(\alpha_1,x).$$

The DAs can be described in a systematic way using conformal expansion
\cite{BBKT};
here, we restrict ourselves to the leading terms in that expansion and
use the expressions \cite{prep} ($\alpha_3=1-\alpha_1-\alpha_2$)
\begin{eqnarray}
\Phi_{3;K^*}^\parallel(\underline{\alpha})
& = & 360\alpha_1\alpha_2\alpha_3^2 \left\{
     \kappa_{3K}^\parallel + \omega_{3K}^\parallel (\alpha_1-\alpha_2) +
     \lambda_{3K}^\parallel \frac{1}{2}\,(7\alpha_3 -
      3)\right\},
\nonumber\\
\widetilde\Phi_{3;K^*}^\parallel(\underline{\alpha})
& = & 360\alpha_1\alpha_2\alpha_3^2 \left\{
      \zeta_{3K}^\parallel + \widetilde\lambda_{3K}^\parallel 
      (\alpha_1-\alpha_2) + \widetilde\omega_{3K}^\parallel
     \frac{1}{2}\,(7\alpha_3 - 3)\right\},
\nonumber\\
\Phi_{3;K^*}^\perp(\underline{\alpha})
& = & 360\alpha_1\alpha_2\alpha_3^2 \left\{
          \kappa_{3K}^\perp + \omega_{3K}^\perp (\alpha_1-\alpha_2) +
          \lambda_{3K}^\perp \frac{1}{2}\,(7\alpha_3 -
           3)\right\}\,,
\nonumber\\
\Phi_{4;K^*}^{\perp(1)}(\underline{\alpha})
 & = & 
120\alpha_1\alpha_2\alpha_3 \left(\frac{1}{4}\,\kappa_{3K}^\perp +
\frac{1}{2}\, \kappa_{4K}^\perp\right),
\nonumber\\
{\Phi}^{\perp(2)}_{4;K^*}(\underline{\alpha})  
&=&  -30 \alpha_3^2\left\{(1-\alpha_3)\left(-\frac{1}{4}\,\kappa_{3K}^\perp +
\frac{1}{2}\, \kappa_{4K}^\perp\right) - (\alpha_1-\alpha_2)
\widetilde\zeta^\perp_{4K} \right\},
\nonumber\\
\Phi^{\perp(3)}_{4;K^*}(\underline{\alpha})
& = & 
-120 \alpha_1\alpha_2\alpha_3 \left(\frac{1}{4}\,\kappa_{3K}^\perp -
\frac{1}{2}\, \kappa_{4K}^\perp\right),
\nonumber\\
{\Phi}^{\perp(4)}_{4;K^*}(\underline{\alpha})  
&=&  30 \alpha_3^2\left\{(1-\alpha_3)\left(-\frac{1}{4}\,\kappa_{3K}^\perp -
\frac{1}{2}\, \kappa_{4K}^\perp\right) - (\alpha_1-\alpha_2)
\zeta^\perp_{4K} \right\},
\nonumber\\
{\Psi}_{4;K^*}^\perp(\underline{\alpha}) 
&=&  30 \alpha_3^2 \left\{(1-\alpha_3) \zeta^\perp_{4K}+
(\alpha_1-\alpha_2) \left( \frac{1}{4} \kappa_{3K}^\perp + 
\frac{1}{2}\kappa_{4K}^\perp\right)\right\},
\nonumber\\
\widetilde{\Psi}_{4;K^*}^\perp(\underline{\alpha}) 
&= &
 30 \alpha_3^2 \left\{(1-\alpha_3)\widetilde\zeta^\perp_{4K}-
(\alpha_1-\alpha_2) \left(- \frac{1}{4} \kappa_{3K}^\perp + 
\frac{1}{2}\kappa_{4K}^\perp\right)\right\}.
\end{eqnarray}
Preliminary numerical results for the various hadronic parameters 
$\zeta$, $\kappa$, $\omega$ and $\lambda$ are
collected in Tab.~\ref{tab:kappas}; they will be discussed in more
detail in Ref.~\cite{prep}.
\begin{table}[tb]
\renewcommand{\arraystretch}{1.3}
\addtolength{\arraycolsep}{3pt}
$$
\begin{array}{l||c|l}
& \mu= 1\,{\rm GeV} & \mbox{Remarks}\\\hline
\zeta_{3K}^\parallel & 0.033 \pm 0.007  & \mbox{new; $\zeta_{3\rho}^\parallel$
       determined in \cite{ZZC}}\\
\widetilde\lambda_{3K}^\parallel & 0.06\pm 0.03 & \mbox{G-odd, new}\\
\widetilde\omega_{3K}^\parallel & -0.06\pm 0.02 &
   \mbox{new; $\widetilde\omega_{3\rho}^\parallel$ determined in \cite{ZZC}}\\
\kappa_{3K}^\parallel & 0.001\pm 0.001 & \mbox{G-odd; previously determined
  in \cite{BZ06_2}}\\
\omega_{3K}^\parallel & 0.14\pm 0.03 &
\mbox{new; $\omega_{3\rho}^\parallel$  determined in
  \cite{ZZC}}\\
\lambda_{3K}^\parallel & -0.02\pm 0.01 & \mbox{G-odd, new}\\
\kappa_{3K}^\perp & 0.006\pm 0.003 & \mbox{G-odd, new}\\
\omega_{3K}^\perp & 0.4\pm 0.1 & \mbox{new; 
   $\omega_{3\rho}^\perp$ determined in
  \cite{BBKT}}\\
\lambda_{3K}^\perp & -0.05\pm 0.02 & \mbox{G-odd, new}\\
\zeta_{4K}^\perp & 0.10\pm 0.05 & \mbox{quoted from \cite{BBK};
  no SU(3) breaking; to be updated in \cite{prep}}\\
\widetilde\zeta_{4K}^\perp & =-\zeta_{4K}^\perp & \mbox{quoted 
  from \cite{BBK};
  no SU(3) breaking; to be updated in \cite{prep}}\\
\kappa_{4K}^\perp & 0.012\pm 0.004 & \mbox{G-odd; quoted from \cite{BZ06_1}}
\end{array}
$$
\renewcommand{\arraystretch}{1}
\addtolength{\arraycolsep}{-3pt}
\vspace*{-20pt}
\caption[]{\small Three-particle twist-3 and 4 hadronic parameters. All 
  results labelled ``new'' are preliminary and will be finalised 
in Ref.~\cite{prep}. Note that the absolute sign of all these parameters
  depends on the sign convention chosen for the strong coupling
  $g$. The above results correspond to the choice $D_\mu =
  \partial_\mu - i g A^a_{\mu} (\lambda^a/2)$ of the covariant
  derivative.}\label{tab:kappas}
\end{table}
The DAs defined above are related to those introduced in
Ref.~\cite{BBKT,rho} as
\begin{equation}
\begin{array}[b]{l@{\quad}l@{\quad}l}
\Phi^{\parallel}_{3;K^*} = {\cal V}\,,&
\widetilde\Phi^{\parallel}_{3;K^*} = {\cal A}\,,&
\Phi^\perp_{3;K^*} = {\cal T}\,,\\[5pt]
\Phi^{\perp(i)}_{4;K^*} = T^{(i)}\,,&
\Psi^\perp_{4;K^*} = S\,,&
\widetilde\Psi^\perp_{4;K^*} = \widetilde S\,.
\end{array}
\end{equation}
Although the introduction of new notations may, at first, look
unmotivated, it actually extends the labelling scheme
introduced, in Ref.~\cite{BBL06}, for pseudoscalar mesons, to vector
mesons and aims to provide a systematic
way to label the multitude of two- and three-particle
pseudoscalar and vector meson DAs, replacing
the slightly ad-hoc notations introduced in our previous papers on
the subject \cite{BBKT,rho}.   
As for the other hadronic parameters entering (\ref{1}) and (\ref{2}), we use  
$m_b = (4.7\pm 0.1)\,{\rm GeV}$, $f_B = (200\pm 30)\,$MeV, 
$f_{K^*}^\parallel = (217\pm 5)\,$MeV \cite{PDG} and
$f_{K^*}^\perp(1\,{\rm GeV}) = (185\pm 10)\,$MeV \cite{BZ05}. All
scale-dependent parameters are evaluated at the scale $\mu^2 =
m_B^2-m_b^2\pm 1\,{\rm GeV}^2$, see Ref.~\cite{BZ04}. $s_0$
and $M^2$ are sum rule specific parameters which do not acquire sharp
values, but have to be varied in a certain range. Based on our
experience with $B$ decay form factors \cite{BZ04} we choose 
$s_0 = (35\pm 2)\,{\rm GeV}^2$ and $M^2=(10\pm 3)\,{\rm GeV}^2$. 
We then obtain
\begin{equation}
L = (0.2\pm 0.1)\,{\rm GeV}^3\,, \quad \widetilde L = (0.3\pm 0.2)
\,{\rm GeV}^3\,,
\quad L-\widetilde L = -(0.1\pm 0.1)\,{\rm GeV}^3\,.\label{21}
\end{equation}
It turns out that the contribution of the $(m_{K^*}/m_b)^2$ terms to
the sum rules is tiny, so that the result and its uncertainty is
entirely dominated by $m_b$, $f_B$ and 
the twist-3 DAs $\Phi^\parallel_{3;K^*}$ and
$\widetilde\Phi^\parallel_{3;K^*}$. We repeat that the parameters
describing these DAs, collected in Tab.~\ref{tab:kappas}, are preliminary.

The results in (\ref{21}) refer to the renormalisation scale
$\mu^2=m_B^2-m_b^2\approx(2.2\,{\rm GeV})^2$. 
Unfortunately, the dependence of $L$ and
$\widetilde L$ on $\mu$ is unknown. We can, however, estimate the
potential impact of a change of scale by evaluating the light-cone sum
rules at the higher scale $\mu=m_b$, although this is, strictly
speaking, incorrect 
in that framework. Nonetheless, $L$ and
$\tilde L$ itself decrease by about 20\% by this procedure, whereas
$L-\widetilde L$ decreases by 10\%, which is well within the
quoted errors. 

Comparing with the results obtained in Ref.~\cite{alex}, Eq.~(\ref{alex}),
we find that our central values are considerably smaller. As mentioned
before, the authors of \cite{alex} used local sum rules, which are of
only limited value for determining $B$ decay form factors at maximum
recoil, i.e.\ for maximum energy of the final state meson, see the
discussion in the first two references in \cite{studies}. On the other
hand, in 1997 not much was known about three-particle twist-3 DAs of
vector mesons, so local QCD sum rules were the best tool at hand at
the time. We also find
that our errors are larger than those in (\ref{alex}), 
which is due to the fact that the
uncertainties quoted in (\ref{alex}) are obtained by varying only the sum
rule parameters $s_0$ and $M^2$, but not the hadronic input parameters.

\section{Results and Conclusions}

We are now finally ready to present results for the CP asymmetry $S$
in (\ref{-1}). The $m_s$-dependent terms in (\ref{0}) yield
\begin{equation}\label{res1}
S^{{\rm SM},s_R} = -0.027 \pm 0.006(m_{s,b}) \pm 0.001
(\sin(2\beta))\,,
\end{equation}
where we use $m_s(2\,{\rm GeV}) = (100\pm 20)\,$MeV \cite{ms},
$m_b(m_b) = (4.20\pm 0.04)\,$GeV \cite{mb} and $\sin(2\beta) =
0.685\pm 0.032$ \cite{HFAG}. One can estimate the impact of radiative
corrections on that result by comparing it with the perturbative QCD
calculation of Ref.~\cite{sanda}. The authors of \cite{sanda} 
obtain the same central value
for $S^{{\rm SM},s_R}$ and also quote, very helpfully, results obtained for
neglecting various sources of corrections, in particular $
-0.034\pm 0.013$ if all long-distance contributions are
neglected. From this we conclude that the impact of radiative
corrections on (\ref{res1}) is likely to slightly increase the
asymmetry, but not by more than 0.01. As for the contribution of
$L-\tilde L$, it is given, to leading order in $\alpha_s$, by
\begin{equation}\label{res2}
S^{\rm SM, soft~gluons} = -2 \sin(2\beta) \left( - \frac{C_2}{C_7}
\,\frac{L-\tilde L}{36m_b m_c^2 T_1^{B\to K^*}(0)}\right) = 0.005\pm 0.01\,.
\end{equation}
Here we use $C_2(m_b) =1.02$, $C_7(m_b)= -0.31$, which are the
leading-order values, $m_c=1.3\,$GeV
and $T_1^{B\to K^*}(0)=0.31\pm0.04$ \cite{proc},
and have doubled the error to account for
neglected higher-order terms in the $1/m_c$ expansion. 
That is: the contribution of soft gluons to $S$ is  much smaller numerically
than that in $m_s/m_b$, Eq.~(\ref{res1}). This
result has to be compared with the dimensional estimate presented in
Ref.~\cite{grin05}, from a SCET-based analysis,
\begin{equation}\label{res3}
|S^{\rm SM,  soft~gluons}_{\mbox{\scriptsize
\cite{grin05}}}| = 2 \sin(2\beta)\,
\left|\frac{C_2}{3C_7}\right|\frac{\Lambda_{\rm QCD}}{m_b}\approx 0.06\,.
\end{equation}
Our result (\ref{res2}) suggests that the true value of the soft gluon 
contributions is much smaller. Comparing (\ref{res2}) and (\ref{res3}),
it becomes obvious that this is mainly due to the factor $1/36$ in
(\ref{res2}) resulting from the short-distance expansion of the charm loop
in Fig.~\ref{fig1}.

While the aim of our letter was to calculate the soft gluon
contributions to $S^{\rm SM}$ induced by the
operator $Q_2$ and to check the estimate of
Ref.~\cite{grin05} that it could induce a 10\% effect, our result for 
$S^{\rm SM,  soft~gluons}$ has now become, due to the suppression
factor $1/36$, that small that one may start to wonder about the size
of other corrections. One source of such corrections, and actually 
probably the dominant one, are radiative corrections to the term in
$m_s/m_b$ which we estimate using the results of
Ref.~\cite{sanda} in the way discussed above. Another class of (soft gluon) corrections are
diagrams with the same topology as Fig.~1, but a different
operator. As long as there is a charm quark in the loop, these
contributions are controlled by the matrix element $L-\tilde L$, but
suppressed by small penguin Wilson-coefficients $C_{\rm peng}< 0.1$ and
hence can be neglected. For light quarks in the loop, one cannot apply
the short-distance expansion as done in this letter, but has to follow
a different approach. We will discuss this approach in a forthcoming
paper on power-corrections to $B\to\rho\gamma$ \cite{prep2}; the result is that
the contribution of light quark loops is of approximately the
same size as that of charm loops, so again these contributions are
suppressed by small Wilson coefficients. A second, different topology
is given by annihilation diagrams induced by the penguin operators $(\bar s
b)_{V-A}(\bar d d)_{V\pm A}$. This contribution is enhanced by the
fact that it is a tree diagram; it can be
calculated using the results obtained in Ref.~\cite{emi} for
$B\to\gamma e\nu$ transitions. For the contribution with the largest
Wilson-coefficient from the penguin operator $Q_4$, one has
$$
a_{7R}^c\to a_{7R}^c + C_4 \,\frac{Q_d}{Q_u} \,\frac{2\pi^2
  f_{K^*}m_{K^*}}{ m_B^2 T_1^{B\to K^*}(0)}\, (F_V(0)-F_A(0))\,,
$$
where $Q_{u,d}$ are the electric charges of the corresponding quarks and 
$F_{V,A}$ are the form factors determining the $B\to \gamma$ transition.
Using $F_V(0)-F_A(0) \approx 0.016$ \cite{emi} and $C_4(m_b) = 0.08$, 
the shift of $a_{7R}^c$ turns out to be $\approx 0.3\cdot 10^{-3}$
which is to be compared with the (dominant) $m_s/m_b$ term $\approx
6\cdot 10^{-3}$ and the term in $C_2$: $\approx 1\cdot 10^{-3}$. Let
us note in passing that $F_V(0)-F_A(0)$ is induced by long-distance
photon emission and given in terms of
three-particle Fock states of the photon \cite{photon}, so also for
this contribution the necessary spin-flip in the parton-level process
$b\to s\gamma$ is induced by a higher Fock state, this time of the
photon.
One more possible topology are hard-spectator scattering diagrams
involving the chromomagnetic dipole operator $Q_8$. Although we cannot
give a firm estimate of this contribution to $a_{7R}^c$, we expect it
to come mainly from long-distance photon emission governed by the same
three-particle Fock state of the photon mentioned before and to
contribute at the same level as the other terms discussed above.
Although we were not able to identify a further sizable contribution 
we will add an uncertainty of $0.01$ to the asymmetry.

Our final result for the CP asymmetry is the sum of (\ref{res1}) and
(\ref{res2}) with uncertainties added in quadrature: 
\begin{equation}\label{res4}
S^{\rm SM} = S^{{\rm SM},s_R} + 
S^{\rm SM, soft~gluons} =  -0.022\pm 0.015 ^{+0}_{-0.01}\,,
\end{equation}
where the second uncertainty accounts for neglected contributions
induced by penguin operators and the chromomagnetic dipole operator,
and the third, asymmetric uncertainty is to account for neglected
$O(\alpha_s)$ corrections to (\ref{res1}), which, based on the results of
Ref.~\cite{sanda}, we estimate to be negative and not to exceed 0.01. 
In principle these corrections can also be calculated in 
QCD factorisation, 
but this goes beyond the scope of this letter.

To summarize, we have calculated the dominant contributions to
the SM prediction for the time-dependent CP asymmetry $S$ in
$B^0\to K^{*0}\gamma$. These come, on the one hand, from terms in
$m_s/m_b$, and on the other hand from short-distance processes
involving an additional gluon, see Fig.~\ref{fig1}. We find that in
contrast to recent suggestions that the latter be large, they are
actually substantially smaller than the former. Additional hadronic corrections
to our result are expected to be even smaller and due to radiative
corrections, small Wilson coefficients and higher order terms in the
heavy quark expansion. The most dominant correction is likely to be
radiative corrections to (\ref{res1}), which have already been
calculated in perturbative QCD and found to be $\approx -0.01$. A
confirmation of this result in QCD factorisation, if possible, would
be welcome. Our result (\ref{res4})  confirms that the CP asymmetry is
an excellent quasi
null test of the SM in the sense of Ref.~\cite{null} and that any
significant deviation of the experimental result from zero will
provide a clean signal for new physics.

\end{document}